\documentclass[a4paper,11pt]{article}
\pdfoutput=1 

\usepackage{jcappub} 

\usepackage[T1]{fontenc} 
\usepackage{color}
\definecolor{Red}{rgb}{0.65,0.08,0.05}
\definecolor{Blue}{rgb}{0.05,0.08,0.65}

\newcommand{\dd}{{\rm d}}
\newcommand{\ii}{{\rm i}}

\newcommand{\hn}{\hat{n}}
\newcommand{\hk}{\hat{k}}
\newcommand{\vx}{\textbf{x}}

\newcommand{\vq}{\textbf{q}}
\newcommand{\vk}{\textbf{k}}

\newcommand{\etain}{{\eta_{\rm in}}}

\newcommand{\mH}{{\cal H}}
\newcommand{\partialeta}{\partial_{\eta}}
\newcommand{\tPsi}{{\tilde\Psi}}

\newcommand{\mF}{{\cal F}}

\newcommand{\vPi}{{\tau}}
\newcommand{\tP}{{P}}
\newcommand{\iniv}{{v}}

\begin{document}
\title{Describing massive neutrinos in cosmology as a collection of independent flows}

\author[a]{H\'el\`ene Dupuy,}
\author[a]{Francis Bernardeau}
\affiliation[a]{Institut de Physique Th\'eorique, CEA, IPhT, F-91191 Gif-sur-Yvette,\\
CNRS, URA 2306, F-91191 Gif-sur-Yvette, France}

\emailAdd{helene.dupuy@cea.fr}      
\emailAdd{francis.bernardeau@cea.fr}

\abstract{
A new analytical approach allowing to account for massive neutrinos in the non--linear description of the growth of the 
large--scale structure of the universe is proposed. Unlike the standard approach in which neutrinos are  described as a unique 
hot fluid, it is shown that the overall neutrino fluid can be equivalently decomposed into a collection of independent flows. Starting either from 
elementary conservation equations or from the evolution equation of the phase--space distribution function, we derive the two non--linear motion equations 
that each of these flows satisfies. Those fluid equations describe the evolution of macroscopic fields. We explain in detail the connection between the collection of flows we defined and the standard massive neutrino fluid. Then, in the particular case of adiabatic  initial conditions,
we explicitly check that, at linear order, the resolution of this new system of equations reproduces the results obtained in the standard approach based on the collisionless Boltzmann hierarchy. Besides, the approach advocated in this paper allows to show how each neutrino flow settles into the cold dark matter flow depending on initial velocities. It opens the way to a fully non--linear treatment of the dynamical evolution of neutrinos in the framework of large--scale structure growth.
}

\date{\today}
\maketitle

\bigskip

\section{Introduction}
The recent results of the Planck mission \cite{PlanckCollaboration2013,2013arXiv1303.5076P} crown three decades of observational and theoretical investigations on the origin, evolution and statistical properties of  cosmological perturbations. Those properties are governed not only by the mechanisms that produced cosmological perturbations -- inflation is the most commonly referred explanation -- but also by matter itself. In addition to the information they give regarding inflationary parameters, observations of the Cosmic Microwave Background (CMB) temperature anisotropies and polarization are thus a very precious probe of the matter content of the universe.
Although this observational window is confined to the time of recombination, in a regime where the metric or density perturbations are all deeply rooted in the linear regime\footnote{Except the effects of lensing, which reveal non--linear line--of--sight effects.}, it allows an exquisite determination of several fundamental cosmological parameters. However, some of them remain elusive. This is in particular the case of the neutrino masses if they are too small to leave an imprint on the recombination physics.

The experiments on neutrino flavor oscillations demonstrating that neutrinos are indeed massive are thus of crucial importance and it is necessary to examine minutely the impact of those masses on various cosmological observables.
Understandably, such a discovery has triggered a considerable effort in theoretical, numerical and observational 
cosmology to infer the consequences on the cosmic structure growth. 
The first study in which massive neutrinos are properly treated in the  linear theory of gravitational perturbations dates back from Ref. \cite{1994ApJ...429...22M} (see also its companion paper Ref. \cite{Ma:1995ey}). The consequences of these
results are thoroughly presented in Ref. \cite{Lesgourgues2006}, where the connection between neutrino masses and cosmology - in the standard case of three neutrino species - is investigated in full detail. It is shown that CMB anisotropies are indirectly sensitive to massive neutrinos whereas the late--time large--scale structure growth rate, via its time and scale dependences, offers a much more direct probe of the neutrino mass spectrum. To a large extent current and future cosmology projects aim at exploiting these dependences to put constraints on the neutrino masses. Indeed, the impact of massive neutrinos on the structure growth has proved to be significative enough to make such constraints possible, as shown for instance in 
\cite{2011PhRvD..83d3529S,2012PhRvD..85h1101R,2013JCAP...01..026A,2011JCAP...03..030C,2011arXiv1110.3193L,2009A&A...500..657T}.
These physical interpretations are based on numerical experiments, the early incarnations dating back from the work of Ref. \cite{1996ApJ...470..102K}, 
which have witnessed a renewed interest in the last years  \cite{2010JCAP...09..014B,2012MNRAS.420.2551B,2012JCAP...02..045H,2013MNRAS.428.3375A}, and also on theoretical investigations such as
\cite{2009JCAP...06..017L,2009PhRvD..80h3528S,2008JCAP...10..035W,2008PhRvL.100s1301S}, 
where the effect of massive neutrinos in the non--linear regime is investigated with the help of Perturbation Theory.
An important point is that it is potentially possible to get better constraints than what the predictions of linear theory offer.
Observations of the large--scale structure within the local universe are indeed sensitive to the non--linear growth of structure and thus also to the impact of mode--coupling effects on this growth. Such a coupling is expected to strengthen the role played by the matter and energy content of the universe on cosmological perturbation properties. This is true for instance for the dark energy equation of state \cite{2001PhRvD..64h3501B} or for the masses, even if small, of the neutrino species, as shown in numerical experiments \cite{2012MNRAS.420.2551B}.

One can mention another alternative that has been proposed to study the effect of neutrinos on the large--scale structure growth, \cite{2010PhRvD..81l3516S}, where the neutrino fluid is tentatively described as a perfect fluid. 
In the present study we are more particularly interested in designing tools to explore the impact of massive neutrinos within the non--linear regime of the density perturbation growth. Little has been obtained in this context in  presence of massive neutrinos. One of the reasons for such a limitation is that the non--linear evolution equations of the neutrino species are a priori cumbersome and difficult to handle (the most thorough investigations of the non--linear hierarchy equations are to be found in \cite{vanderijtphd}). On the other hand Perturbation Theory applied to pure dark matter systems has proved very valuable and robust (see \cite{2013arXiv1311.2724B} for a recent review on the subject).
The aim of this paper is thus to set the stage for further theoretical analyses by presenting a complete set of equations
describing the neutrino perturbation growth, from super--Hubble perturbations of relativistic species to those of non--relativistic species 
within the local universe.  In particular we are interested in deriving equations from which the connection with the standard 
non--linear system describing dark matter particles is convenient.

The strategy usually adopted to describe neutrinos, massive or not, is calqued from that used to describe the radiation fluid
(see e.g. Refs. \cite{Lesgourgues2006,1994ApJ...429...22M,Ma:1995ey}): neutrinos are considered as a single hot multi-stream
fluid whose evolution is dictated by the behavior of its distribution function in phase--space $f$. Calculations are performed in a perturbed Friedmann--Lema\^itre spacetime. The key equation is the Boltzmann equation. For neutrinos, contrary to radiation, it is taken in the collisionless limit since neutrinos do not interact with ordinary matter (neither at the time of recombination nor after).  It leads to the 
 Vlasov equation, which derives from the conservation of the number of particles applied to a Hamiltonian system, $\dfrac{\dd{f}}{\dd{\eta}}=0$, where $\eta$ is a time coordinate. The different terms participating in the expanded form of this equation are computed in particular with the help of the geodesic equation. In practice, whether at super or sub--Hubble scales, the motion equations are derived at linear order with respect to the metric fluctuations. We will here also restrict ourselves to this approximation as the late--time non--linearity of the large--scale structure growth is not due to direct metric--metric couplings but to the non--linear growth of the density contrasts and velocity divergences.

The possibility we explore in the present work is that neutrinos\footnote{As a matter of fact, neutrinos of each mass eigenstate.} could be considered as a collection of single--flow fluids instead of a single multi--flow fluid. We take advantage here of the fact that neutrinos are actually free streaming: they do not interact with one another and they do not interact with  matter particles. We will see in particular that it is possible to distinguish the fluid elements of the collection by labeling each of them with an initial velocity. The complete neutrino fluid behavior is then naturally obtained by summing the contributions of each fluid element over the initial velocity distribution. As we will see, this description is actually very similar to that of dark matter from the very beginning. It also breaks down for the same reason: an initially single--flow fluid can form multiple streams after shell crossings.
But this regime corresponds to the late--time evolution of the fields, which is beyond the scope of Standard Perturbation Theory calculations so no shell--crossing is taken into account in the following.

The paper is organized as follows. In Sect. \ref{EqMotions}, the geometric context in which the calculations are performed is specified as well as some physical quantities of interest. We then derive the non--linear equations of motion associated with each fluid of neutrinos and we make the comparison with the first moments of the Boltzmann hierarchy. In Sect. \ref{Linear}, we describe the linearized system. Sect. \ref{Sect:multiflow}  is devoted to the description of the specific construction of a multi--fluid system of single flows. We present in particular the initial and early--time number density and velocity fields corresponding to adiabatic initial conditions. This section ends by the presentation of the results given by the numerical integration of the system of equations, when the whole neutrino fluid is discretized into a finite sum of independent fluids. 
Results are explicitly compared to those obtained from the standard integration scheme based on the Boltzmann hierarchy.
Finally we give some hints on how each flow settles into the cold dark matter component.

\medskip
   
\section{Equations of motion}
\label{EqMotions}

In this section we present the derivation of the non--linear equations of motion for a single--flow fluid of particles, relativistic or not. A confrontation of our findings with those of a more standard approach based on the use of the Vlasov equation is presented in the last part of the present section.

\subsection {Spacetime geometry, momenta and energy}

In order to describe the impact of massive neutrinos on the evolution of inhomogeneities, we consider a spatially flat Friedmann--Lema\^itre spacetime with scalar metric perturbations only. Units are chosen so that the speed of light in vacuum is equal to unity. We adopt in this work the
conformal Newtonian gauge, which makes the comparison with the standard motion equations of 
non--relativistic species, the Vlasov-Poisson system, easier. The metric is given by
\begin{equation}\label{metric}
\dd{s}^2=a^2\left(\eta\right)\left[-\left(1+2\psi\right)\dd{\eta}^2+\left(1-2\phi\right)\dd{x^i}\dd{x^{j}}\delta_{ij}\right],
\end{equation}
where $\eta$ is the conformal time, $x^i$ ($i =1,2,3$) are the Cartesian spatial comoving coordinates, $a\left(\eta\right)$ is the scale factor , $\delta_{ij}$ is the Kronecker symbol  and $\psi$ and $\phi$ are the metric perturbations.
The expansion history of the universe, encoded in the time dependence of $a$, is driven by the overall
matter and energy content of the universe. It is supposed to be known and for practical calculations we adopt the numerical values of the concordance model.

Following the same idea, in the rest of the paper metric perturbations will be considered as known, determined by the Einstein equations. 
Furthermore, following the framework presented in the introduction, only linear terms in $\psi$ and $\phi$ will be taken into account
in all the derivations that follow, in particular in the motion equations we will derive. 

We will consider massive particles, relativistic or not, freely moving in space--time (\ref{metric}). 
Their kinematic properties
are given by their momenta so we introduce the quadri-vector $p_{\mu}$ as the conjugate momentum of $x^{\mu}$, i.e.
\begin{equation}\label{Pdef}
p_{\mu}=m u_{\mu}\ \ \hbox{where}\ \ u_{\mu}=g_{\mu\nu}\dd x^{\nu}/\sqrt{-\dd s^{2}}.
\end{equation}
It obviously implies $p^{\mu}p_{\mu}=-m^{2}$. 
In the following we will also make use of the momentum $q^{i}$, defined as 
\begin{equation}\label{qidef}
q^{i}=a^2(1-\phi)p^i
\end{equation}
in such a way that
\begin{equation}
p_{i}p^{i}=g_{ij}p^{i}p^{j}=\delta_{ij}\frac{q^{i}q^{j}}{a^{2}}.
\end{equation}
Another useful quantity is the energy $\epsilon$ measured by an observer at rest in metric (\ref{metric}), which is such that
\begin{equation}
p^{0}p_{0}=g_{00}p^{0}p^{0}\equiv -{\epsilon^{2}}.
\end{equation}
It satisfies
\begin{equation}
\epsilon^{2}=m^{2}+(q/a)^{2}.
\end{equation}

\subsection{The single--flow equations from conservation equations}

We now proceed to the derivation of the motion equations satisfied by a single--flow fluid starting from elementary conservation equations. Such a fluid is entirely characterized by two fields, its local numerical density field $n(\eta,\vx)$ and its velocity or momentum\footnote{We use an uppercase to distinguish it from a phase--space variable.} field $P^{i}(\eta,\vx)$ (the zero component can be deduced from the spatial ones using the on--shell
mass constraint). This approach contrasts with a description of the complete neutrino fluid for which one has to introduce the whole velocity distribution.

\subsubsection{Evolution equation of the proper number density $n$}

The key idea is to consider a set of neutrinos that form a single flow, i.e. a fluid in which there is only one velocity (one modulus and one direction) at a given position. If a fluid initially satisfies this condition, it will continue to do so afterwards since the neutrinos it contains evolve in the same gravitational potential.  Thus in the following we consider fluids in which all the neutrinos have initially the same velocity.
In such physical systems, neutrinos are neither created 
nor annihilated nor diffused through collision processes so neutrinos contained in each flow obey an elementary  conservation law,
\begin{equation}\label{Jconservation}
J^\mu_{\phantom{0};\mu}=0,
\end{equation}
where $J^\mu$ is the particle four--current and where we adopt the standard notation ; to indicate a covariant derivative. It is easy to show that, in the metric we chose, this relation leads to,
\begin{align}\label{Jconservation1}
\partialeta{J^0}+\partial_i J^i +( 4 \mathcal {H}+\partialeta {\psi}-3\partialeta {\phi})J^0+(\partial_i \psi -3\partial_i \phi)J^i=0,
\end{align}
where $\mH$ is the conformal Hubble constant, $\mH=\partialeta a/a$.

The four--current is related to the number density of neutrinos as measured by an observer at rest in metric (\ref{metric}), $n(\eta,\vx)$, by  $n=J^\mu U_\mu$, where $U_\mu$ is a vector tangent to the worldline of this observer. The latter satisfies $U^\mu U_\mu=-1$ and $U^i=0$. Thus $n=J^0 U_0 = a\left(1+\psi\right) J^0$. Given that $J^i=J^0 \dfrac{P^i}{P^0}$, Eq.  (\ref{Jconservation1})
can thereby be rewritten 
\begin{equation}\label{n haut}
\partialeta n+\partial_{i}\left(\dfrac{P^i}{P^0} n\right)=3n\left(\partialeta {\phi}-\mathcal{H}+\partial_i \phi \dfrac{P^i}{P^0}\right).
\end{equation}
We signal here that this number density is the proper number density, not to be confused with the comoving number density that we will define later (Eq. (\ref{comoving number density})). The fact that the right--hand side of its evolution equation is non--zero is thus not surprising. It simply reflects the expansion of the universe.
This relation can alternatively be written with the help of the momentum $P_{i}$, expressed with covariant indices,  
thanks to the relations $P_i=a^2\left(1-2\phi\right) P^i$ and $P_0=-a^2 \left(1+2\psi\right)P^0$.  It leads to,
\begin{equation}\label{neom}
\partialeta n-(1+2\phi+2\psi)\partial_{i}\left(\dfrac{P_{i}}{P_0} n\right)=3n(\partialeta {\phi}-\mathcal{H})+n(2\partial_i \psi-\partial_i \phi)\dfrac{P_i}{P_0},
\end{equation}
where a summation is still implied on repeated indices. Note that in all these transformations we consistently 
keep all contributions to linear order in the metric perturbations\footnote{We note however that the factor $(1+2\phi+2\psi)$ that appears 
in the second term of this equation could be dropped as it is multiplied by a gradient term that vanishes at homogeneous level. For the sake of consistency we however keep such factor here and in similar situations in the following.}.

\subsubsection{Evolution equation of the momentum $P_i$}
\label{EvolutionPi}

The second motion equation expresses the momentum conservation. It is obtained from the observation that, for a single--flow fluid, all particles located at the same position have the same momentum so that the energy momentum tensor $T^{\mu \nu}$ is given by $T^{\mu \nu}=P^\mu J^\nu$. 
The conservation of this tensor  then gives
\begin{equation}\label{T and J1}
T^{\mu \nu}_{\phantom{00};\nu}=P^{\mu }_{\phantom{0};\nu}J^\nu+P^\mu J^\nu_{\phantom{0};\nu}=0,
\end{equation}
which combined with equation (\ref{Jconservation}) leads to 
\begin{equation}\label{T and J}
T^{\mu \nu}_{\phantom{00};\nu}=P^{\mu }_{\phantom{0};\nu}J^\nu=0. 
\end{equation}

This relation should be valid in particular for spatial indices, $P^{i}_{\phantom{0};\nu}J^\nu=0$. It  eventually imposes,
\begin{equation}\label{Peom}
\partialeta {P_i}-(1+2\phi+2\psi)\dfrac{P_j }{P_0}\partial_j P_i=P_0 \partial_i \psi+\dfrac{P_j P_j}{P_0}\partial_i \phi
\end{equation}
on the covariant coordinates of the momentum. Eq. (\ref{Peom}) is our second non--linear equation of motion.
At this stage the fact that we choose covariant coordinates $P_{i}$ instead 
of contravariant $P^{i}$ or a combination of both such as $q_{i}$ is arbitrary but we will see that it is crucial when it comes 
to actually solve  this system in the linear regime, see Sect. \ref{Linear}. 

We have now completed the derivation of our system of equations.
It is a generalization of the standard single--flow equations of a pressureless fluid composed
of non--relativistic particles. The latter is obtained simply by imposing to the velocity to be small compared to unity (while keeping its
gradient large). To see it more easily, let us express the motion equations in terms of the proper
velocity field.  

\subsubsection{Other formulations of the momentum conservation}

An alternative representation of Eq. (\ref{Peom}) can be obtained by introducing the physical velocity {field} $V^{i}(\eta,\vx)$ (expressed 
in units of the speed of light). This velocity is along the momentum $P^{i}$ and is such that $V^{2}=-P_{i}P^{i}/(P_{0}P^{0})$.
We can easily show that,
\begin{equation}
V^{i}=-\frac{P_{i}}{P_{0}} (1+\phi+\psi).
\end{equation}
Note that $V^{i}$ can entirely be expressed in terms of $P_{i}$ with the help of the relation
\begin{equation}
P_{0}^{2}=P_{i}^{2}(1+2\phi+2\psi)+m^{2}a^{2}(1+2\psi).
\end{equation}
Its evolution equation derives from Eq. (\ref{Peom}) and reads
\begin{equation}\label{evolvi}
\partialeta V^{i}+
V^{i}(\mH-\partialeta\phi)(1-V^{2})+\partial_{i}\psi+V^{2}\partial_{i}\phi
+(1+\phi+\psi)V^{j}\partial_{j}V^{i}-V^{i}V^{j}\partial_{j}(\phi+\psi)=0.
\end{equation}
From this equation, it is straightforward to recover the standard Euler equation in the limit of non--relativistic particles.

Similarly, the evolution equation of the energy field $\epsilon(\eta,\vx)$, defined by\footnote{$P_0=-a(1+\psi)\epsilon$ is a sign convention that we use in all this paper.}
\begin{equation}
\epsilon(\eta,\vx)=\frac{m}{\sqrt{1-V(\eta,\vx)^{2}}}=-\dfrac{(1-\psi)}{a}P_0(\eta,\vx),
\end{equation}
can be deduced from Eq. (\ref{Peom}),
\begin{equation}
\label{epsilonevol}
\partialeta \epsilon+(1+\phi+\psi)V^i\partial_{i}\epsilon+\epsilon V^i\partial_{i}\psi +\epsilon V^{2}(\mH-\partialeta \phi)=0.
\end{equation}
We will now compare these field equations to those obtained from the Boltzmann approach, which is based on the evolution
equation of the phase--space distribution function.

\subsection{The single--flow equations from the evolution of the phase--space distribution function}

\subsubsection{The non--linear moments of the Boltzmann equation}\label{NLBA}

The Boltzmann approach consists in studying the evolution of the phase--space distribution function $f(\eta,x^{i},p_{i})$,
defined as the number of particles per differential volume $\dd^{3}x^{i}\dd^{3}p_{i}$ of the phase--space, with respect to the  conformal time $\eta$,  the comoving positions $x^{i}$ and the conjugate  momenta $p_{i}$. The particle conservation implies that
\begin{equation}
\frac{\partial }{\partial\eta}f+
\partial_{i}\left(\frac{\dd x^{i}}{\dd\eta}f
\right)+
\frac{\partial}{\partial p_{i}}\left(\frac{\dd p_{i}}{\dd\eta}f \right)=0,
\label{fdensityevolution}
\end{equation}
where ${\dd x^{i}}/{\dd\eta}$ and ${\dd p_{i}}/{\dd\eta}$ are a priori space and momentum dependent functions.
Because of the Hamiltonian evolution of the system, Eq. (\ref{fdensityevolution}) can be simplified into,
\begin{equation}
\frac{\partial }{\partial\eta}f+
\frac{\dd x^{i}}{\dd\eta}\partial_{i}f+
\frac{\dd p_{i}}{\dd\eta}\frac{\partial}{\partial p_{i}}f=0.
\label{fdensityevolution2}
\end{equation}
In other words, $f$ satisfies a Liouville equation, $\dd f/\dd \eta$=0. To compute 
this total derivative, there is some freedom about the choice of the momentum variable (but this choice does not affect the physical interpretation of $f$, which remains in any case the number of particles per $\dd^{3}x^{i}\dd^{3}p_{i}$). On the basis of previous work, we adopt here the variable $q^{i}$ defined in Eq. (\ref{qidef}). In this context, the chain rule gives for the Liouville equation,
\begin{equation}\label{Vlasov}
\frac{\partial f}{\partial \eta}+\frac{\dd x^{i}}{\dd \eta}\frac{\partial f}{\partial x^{i}}+
\frac{\dd q^{i}}{\dd \eta}\frac{\partial f}{\partial q^{i}}=0.
\end{equation}
We are not interested in deriving a multipole hierarchy at this stage so we keep here a Cartesian coordinate description. From the very definition of $q^{i}$ we have,
\begin{equation}\label{deriv_xi}
\frac{\dd x^{i}}{\dd \eta}=\frac{p^{i}}{p^{0}}=\frac{q^{i}}{a\epsilon}(1+\phi+\psi).
\end{equation}
On the other hand the geodesic equation leads to,
\begin{equation}\label{deriv_pi}
\frac{\dd q^{i}}{\dd \eta}=-a\epsilon\,\partial_i\psi+q^{i}\partialeta \phi+(\hn^{i}\hn^{j}-\delta^{ij})\frac{q^{2}}{a\epsilon}\partial_j\phi,
\end{equation}
where $q^2=\delta_{ij}q^{i}q^{j}$ and $\hn^{i}$ is the unit vector along the direction $q^{i}$,
\begin{equation}
\hn^{i}=\frac{q^{i}}{q}.
\end{equation}
The resulting Vlasov (or Liouville) equation takes the form,
\begin{align}\label{Boltzmann}
\frac{\partial f}{\partial \eta}+&\left(1+\phi+\psi\right)\dfrac{q^i}{a\epsilon}\partial_i f
+a\epsilon \dfrac{\partial f}{\partial q^i}\left[-\partial_i \psi + \dfrac{q^i}{a\epsilon}\partialeta {\phi} +\left(\hat{n^i} \hat{n^j}-\delta^{ij}\right)\dfrac{q^2}{a^{2}\epsilon^2}\partial_j \phi\right]=0.
\end{align}

By definition, the proper energy density $\rho$ and the energy--momentum tensor  are related by $\rho=-T^{0}_{0}$. As demonstrated in \cite{Ma:1995ey}, $\rho$ can thereby be expressed in terms of the distribution function,
\begin{equation}\label{rhodef}
\rho(\eta,\vx)=\int{\dd^3 q^{i}\,\dfrac{\epsilon f}{a^3}}.
\end{equation}
Its evolution equation is obtained by integrating equation (\ref{Boltzmann}) with respect to $\dd^3 q^{i}$ with proper weight,
\begin{equation}\label{Aevol}
\partialeta  \rho+(\mH-\partialeta \phi)(3\rho+A^{ii})+(1+\phi+\psi)\partial_{i}A^{i}+2A^{i}\partial_{i}(\psi-\phi)=0,
\end{equation}
where the quantities $A^{ij...k}$ are defined as,
\begin{eqnarray}
\label{A definition}
A^{ij...k}(\eta,\vx)\equiv\int{\dd^3 q^{i}\left[\dfrac{q^i}{a\epsilon}\dfrac{q^j}{a\epsilon} ... \dfrac{q^k}{a\epsilon}\right]\dfrac{\epsilon f}{a^3}}.
\end{eqnarray}
As explicitly shown in appendix \ref{Boltzmann hier}, a complete hierarchy giving the evolution equations of $A^{ij...k}$ can be obtained following the same idea.

\subsubsection{The single--flow equations from the moments of the Boltzmann equation}

The aim of this paragraph is to show that the motion equations we derived previously, (\ref{neom}) and (\ref{Peom}), can alternatively
be obtained from the Vlasov equation (\ref{Boltzmann}).
This comparison requires to precise the physical meaning of the quantities defined in both approaches in order to explicitly relate them. 

Let us start with the number density of particles. By definition, for any fluid, the \emph{comoving} number density $n_{c}(\eta,\vx)$, i.e. the number of particles per comoving unit volume $\dd^{3}x^{i}$, is related to the distribution function $f$ associated with this fluid thanks to 
\begin{equation}\label{comoving number density}
n_{c}(\eta,\vx)=
\int\dd^{3}p_{i}\ f(\eta,x^{i},p_{i}).
\end{equation}
On the other hand, the \emph{proper} number density $n(\eta,\vx)$, which is such that the proper energy density is given
by $\rho(\eta,\vx)=n(\eta,\vx)\epsilon(\eta,\vx)$, reads
\begin{equation}
n(\eta,\vx)=
\int\dd^{3}q^{i}\ \frac{f(\eta,x^{i},p_{i})}{a^{3}}
\end{equation}
to be in agreement with Eq. (\ref{rhodef}).
Given that $\dd^{3}q^{i}=(1+3\phi)\dd^{3}p_{i}$, the relation between $n_{c}(\eta,\vx)$ and $n(\eta,\vx)$ is therefore
\begin{equation}
n(\eta,\vx)=\frac{1+3\phi(\eta,\vx)}{a^{3}}n_{c}(\eta,\vx)\label{nncrelation}.
\end{equation}
Similarly, the momentum field $P_{i}(\eta,\vx)$ can be defined as the average of the phase--space comoving momenta $p_i$. Using the distribution function to compute this mean value, one thus has
\begin{equation}\label{Pi field}
P_{i}(\eta,\vx)n_{c}(\eta,\vx)=
\int\dd^{3}p_{i}\ {f(\eta,x^{i},p_{i})}p_{i}\ \ \hbox{or}\ \ 
P_{i}(\eta,\vx)n(\eta,\vx)=
\int\dd^{3}q^{i}\ \frac{f(\eta,x^{i},p_{i})}{a^{3}}p_{i}.
\end{equation}

In the particular case explored in this section, fluids are single flows so, for each of them, 
\begin{equation}\label{nf relation}
f(\eta,x^{i},p_{i})=f^{\rm{one-flow}}(\eta,x^{i},p_{i})=n_{c}(\eta,\vx) \delta_{\rm{D}}(p_{i}-P_{i}(\eta,\vx)),
\end{equation}
where $\delta_{\rm{D}}$ is the Dirac distribution function. As a result we have, for any macroscopic field depending on $P_{i}(\eta,\vx)$, $\mF[P_{i}(\eta,\vx)]$,
\begin{equation}\label{any field}
\mF\left[P_{i}(\eta,\vx)\right]n_{c}(\eta,\vx)=\int\dd^{3}p_{i}\ {f(\eta,x^{i},p_{i})}\ \mF\left[p_{i}\right].
\end{equation}
Taking advantage of this, we proceed to show that  the Vlasov and geodesic equations together with the relations (\ref{comoving number density}), (\ref{Pi field}) and (\ref{any field}) allow to recover the equations of motion we derived previously. We first note that the
field $P_{0}(\eta,\vx)$ defined previously as a function of $P_{i}(\eta,\vx)$ is nothing but 
\begin{equation}
P_{0}(\eta,\vx)n(\eta,\vx)=
\int\dd^{3}q^{i}\ \frac{f(\eta,x^{i},p_{i})}{a^{3}}p_{0}.
\end{equation}
It is then easy to see that integration over $\dd^{3}p_{i}$ of the equation (\ref{fdensityevolution})
gives
\begin{equation}
\partialeta n_{c}+\partial_{i}\left(\dfrac{P^i}{P^0} n_{c}\right)=0,
\end{equation}
which is exactly the first equation of motion, (\ref{n haut}), after $n_{c}$ is expressed in terms of $n$ following Eq. (\ref{nncrelation}). Finally, it is also straightforward to show that the average (as defined by Eq. (\ref{any field})) of the geodesic equation (\ref{deriv_pi}) directly gives the second equation of motion when expressed in terms of $P_i$, Eq. (\ref{Peom}).

Note that conversely, it is possible to derive the hierarchy (\ref{Aevol2}-\ref{Aijkevol}) from our equations of motion.
For instance, the combination of  Eqs. (\ref{n haut}) and (\ref{epsilonevol}) gives the following evolution equation for  $\rho(\eta,\vx)=n(\eta,\vx)\epsilon(\eta,\vx)$\footnote{The energy field $\epsilon(\eta,\vx)$ satisfies
$
\epsilon(\eta,\vx) n(\eta,\vx)=\int\dd^{3}q^{i}\ \frac{f(\eta,\vx,\vq)}{a^{3}}\epsilon(q).
$}
\begin{equation}\label{rho eom}
\partialeta\rho+\rho(\mH-\partialeta\phi)(3+ V^{2})+(1+\phi+\psi)\partial_{i}(\rho V^{i})
+2\rho V^{i}\partial_{i}(\psi-\phi)
=0.
\end{equation}
Given that $A(\eta,\vx)=\rho(\eta,\vx)$ and that, for a single--flow fluid, the fields $A^{i_1...i_n}(\eta,\vx)$ are related to $A(\eta,\vx)$ by
\begin{equation}\label{Aisingle}
A^{i}(\eta,\vx)=V^{i}(\eta,\vx)\,A(\eta,\vx),\ \ A^{ij}(\eta,\vx)=V^{i}(\eta,\vx)V^{j}(\eta,\vx)\,A(\eta,\vx),\ \ \hbox{etc.},
\end{equation}
Eq. (\ref{rho eom}) is exactly the average of Eq. (\ref{Aevol}), i.e. Eq. (\ref{Aevol}) multiplied by $f/a^3$, integrated over $\dd^{3}q^{i}$ and divided by $n$. It is then a simple exercise to check that the subsequent equations of the hierarchy can be similarly recovered with successive uses of Eq. (\ref{evolvi}). 

\section{The single--flow equations in the linear regime}\label{Linear}

In this section we explore the system of motion equations (\ref{neom})-(\ref{Peom}) in the linear regime.
 It is useful in particular in order to properly set the initial conditions required to solve the system. 

\subsection{The zeroth order  behavior}

Let us start with the homogeneous quantities. It is straightforward to see that the zeroth order contribution of (\ref{neom}) is 
\begin{equation}
\partialeta {n^{(0)}}=-3\mathcal{H} n^{(0)},
\end{equation}
and that the unperturbed equation for $P_i$ is (see Eq. (\ref{Peom})),
\begin{equation}
\partialeta {P_i}^{(0)}=0.
\end{equation}
As a result the number density of particles is simply decreasing as $1/a^{3}$ and ${P_i}^{(0)}$ is constant.
This latter result is attractive as it makes $P_{i}^{(0)}$ a good variable to label each flow. To take advantage of this property, we introduce a new variable, $\vPi_i$, defined as

 \begin{equation}\label{Pi}
\vPi_{i}\equiv {P_i}^{(0)} \left(\eta\right)={P_i}^{(0)} \left(\eta_{\rm{in}}\right),
\end{equation}
where $\eta_{\rm{in}}$ is the initial time.
We also introduce the norm of $\vPi_{i}$, $\vPi$, given by
\begin{equation}
\vPi=\sqrt{\delta_{ij}\vPi_{i}\vPi_{j}}.
\end{equation}
Similarly, we define $\vPi_{0}$ as ${P_0}^{(0)}$. Note that this quantity is not constant over time. Given the sign convention for $P_0$ adopted in this paper, it satisfies
\begin{equation}
\vPi_{0}=-\sqrt{\vPi^{2}+m^2a^{2}}.
\end{equation}

By definition, $P_i$ is the comoving momentum so the fact that ${P_i}^{(0)}$ is constant does not conflict with the fact that neutrinos, as any other massive particles, tend to ``forget'' their initial velocities and to align with the Hubble flow.  
At this stage, we can already note that\footnote{The behavior of the momentum variables with respect to the scale factor can also be deduced from the geodesic equation (see e.g. Ref.\cite{2013neco.book.....L}). 
\medskip},
\begin{align}
{P_i}^{(0)} \sim \text{constant } \text{  and  }\ \ {P^i}^{(0)}\sim a^{-2}.
\end{align}
To be more comprehensive regarding notations, let us mention that 
the flows can alternatively be labeled by the zeroth order velocity, denoted $\iniv^{i}$,

\begin{equation}
\iniv^i={V^i}^{(0)}.
\end{equation}
It satisfies
\begin{equation}\label{videf}
\iniv^{i}=\frac{\vPi_{i}}{\left(m^2a^{2}+{\vPi^{2}}\right)^{1/2}}=-\frac{\vPi_{i}}{\vPi_{0}},\ \ \iniv^{2}=\delta_{ij}\iniv^{i}\iniv^{j}
\end{equation}
or alternatively,
\begin{equation}
\frac{\vPi_{i}}{a}=\frac{m \iniv^{i}}{\sqrt{1-\iniv^{2}}}.
\end{equation}

\subsection{The first order behavior}

We focus now on the first order system.
 In order to simplify the notations, we introduce the total first order derivative operator as,
\begin{equation}
\dfrac{\dd{{X}^{(1)}}}{\dd{\eta}}=\partial_\eta {X}^{(1)}+\dfrac{{P^i}^{(0)}}{{P^0}^{(0)}}\partial_i {X}^{(1)}.
\end{equation}
It can be written alternatively,
\begin{equation}
\dfrac{\dd{{X}^{(1)}}}{\dd{\eta}}=\partial_\eta {X}^{(1)}-\dfrac{\vPi_i}{\vPi_0}\partial_i {X}^{(1)}=
\partial_\eta {X}^{(1)}+\iniv^i\partial_i {X}^{(1)}.
\end{equation} 
With this notation the first order equation of the number density reads,
 \begin{align}\label{n1}
\dfrac{\dd{{n}^{(1)}}}{\dd{\eta}}=&3\partialeta {\phi} n^{(0)}-3\mathcal{H}{n}^{(1)}+\left(2\partial_i \psi-\partial_i \phi\right)
\dfrac{\vPi_{i}}{\vPi_{0}}
n^{(0)}+\left(\dfrac{\partial_i {P_i}^{(1)}}{\vPi_{0}}-\dfrac{\vPi_{i}\partial_i{P_0}^{(1)}}{\vPi_{0}^{2}}\right)n^{(0)}
\end{align}
and that for the momentum is given by,
\begin{equation}\label{P1}
\dfrac{\dd{{P_i}^{(1)}}}{\dd{\eta}}=\vPi_{0}\partial_i \psi+\dfrac{\vPi^{2}}{\vPi_{0}}\partial_i \phi.
\end{equation}
The latter equation exhibits a crucial property: it shows that the source terms of the evolution of
$P_{i}^{(1)}$ form a gradient field. As a consequence, although one cannot mathematically exclude the existence 
of a curl mode in $P_{i}^{(1)}$, such a mode is expected to be diluted by the expansion so that $P_{i}^{(1)}$ remains effectively
potential. At linear order, this property should be rigorously exact for adiabatic initial conditions\footnote{But there is no guarantee
it remains true to all orders in Perturbation Theory. }.

As a consequence, the $P_{i}^{(1)}$ behavior is dictated by the gradient of $\vPi_0 \psi+\vPi^2/{\vPi_0}\phi$. 
It is not the case of other variables such as ${P^{i}}^{(1)}$, which is a combination of $P_{i}^{(1)}$ and $\iniv^{i}$.  This is the reason we preferably write the motion equations in terms 
of this variable.

We close the system thanks to
the on--shell normalization condition of $P_{\mu}$, which gives the expression of $P_{0}$ at first order,
\begin{align}\label{P01}
P_0^{(1)}
=\dfrac{\vPi_{i}P_{i}^{(1)}}{\vPi_{0}}+\dfrac{\vPi^2}{\vPi_0}\,\phi +\vPi_0\,\psi .
\end{align}
Eqs. (\ref{n1}) and (\ref{P1}) associated with relation (\ref{P01}) form a closed set of equations describing
the first order evolution of a fluid of relativistic or non--relativistic particles.
\subsection{The system in Fourier space}

To explore the properties of the solution of the system (\ref{n1})-(\ref{P1})-(\ref{P01}), let us move to Fourier space.
Each field is decomposed into Fourier modes using the following convention for the Fourier transform,
\begin{equation}
F(\vx)=\int\dfrac{\dd^{3}\vk}{(2\pi)^{3/2}}F(\vk)\exp(\ii\vk.\vx).
\end{equation}
We will consider the Fourier transforms of the density contrast field $\delta_n(\vx)$
\begin{equation}
\delta_{n}(\vx)=\frac{1}{n^{(0)}}n^{(1)}(\vx),
\end{equation}
of the divergence field,
\begin{equation}
\theta_{\tP}(\vx)=\partial_{i}P_{i}^{(1)},
\end{equation}
and of the potentials. 
We can here take full advantage 
of the fact that $P_{i}$ is potential at linear order. It indeed implies that ${P_{i}}^{(1)}$ is entirely characterized by its divergence,
\begin{equation}
P_{i}^{(1)}(\vk)=\frac{-\ii k_{i}}{k^{2}}\theta_{\tP}(\vk).
\end{equation}
After replacing equation (\ref{P1}) by its divergence, one finally obtains from equations (\ref{n1}) and (\ref{P1})
\begin{align}\label{nFourier}
\partialeta \delta_{n}=\ii\mu k \frac{\vPi}{\vPi_{0}}\delta_{n}+
3\,\partialeta \phi+\frac{\theta_{\tP}}{\vPi_{0}}\left(1-\frac{\vPi^{2}}{\vPi_{0}^{2}}\mu^{2}\right)
-\ii\mu k\frac{\vPi}{\vPi_{0}}
\left[\left(1+\frac{\vPi^{2}}{\vPi_{0}^{2}}\right)\phi-\psi\right]
\end{align}
and
\begin{equation}\label{thetaFourier}
\partialeta {\theta_{\tP}}=\ii\mu k \frac{\vPi}{\vPi_{0}}{\theta_{\tP}}-\vPi_{0}k ^{2}\psi-\dfrac{\vPi^{2}}{\vPi_{0}}k^{2}\phi,
\end{equation}
where  $\mu$ gives the relative angle between $\vk$ and $\vPi$ or alternatively between $\vk$ and \textbf{$\iniv$},
\begin{equation}
\mu=\dfrac{\mathbf{k}.\mathbf{\vPi}}{k \vPi}=\frac{\vk.\textbf{$\iniv$}}{k\,\iniv}. 
\label{mudef}
\end{equation}
These equations can alternatively be written in terms of the zeroth order physical velocity $\iniv$,
\begin{eqnarray}
\partialeta \delta_{n}&=&-\ii\mu k \iniv\,\delta_{n}+3\,\partialeta \phi-
{\sqrt{1-\iniv^{2}}}\left(1-\iniv^{2}\mu^{2}\right)\frac{ \theta_{\tP}}{ma}
+\ii\mu k \iniv
\left[\left(1+\iniv^{2}\right)\phi-\psi\right],\label{nFourier2}\\
\partialeta \theta_{\tP}&=&-\ii\mu k \iniv\,\theta_{\tP}+\frac{ma}{\sqrt{1-\iniv^{2}}}k ^{2}\left(\iniv^{2}\phi+\psi\right).\label{thetaFourier2}
\end{eqnarray}
This is this system that we encode in practice. 
As we will see, it provides a valid representation of a fluid of initially relativistic species.
In the following, we explicitly show how it can be implemented numerically.

\section{A multi--fluid description of neutrinos}
\label{Sect:multiflow}

In this section, we explain how one can define a collection of flows to describe the whole fluid of neutrinos.
Note that this construction is valid for any given mass eigenstate of the neutrino fluid. If the masses are not degenerate, it should therefore be repeated for each three eigenstates.

\subsection{Specificities of the multi--fluid description}

\begin{figure}[!h]
\begin{center}
\includegraphics[width=10cm]{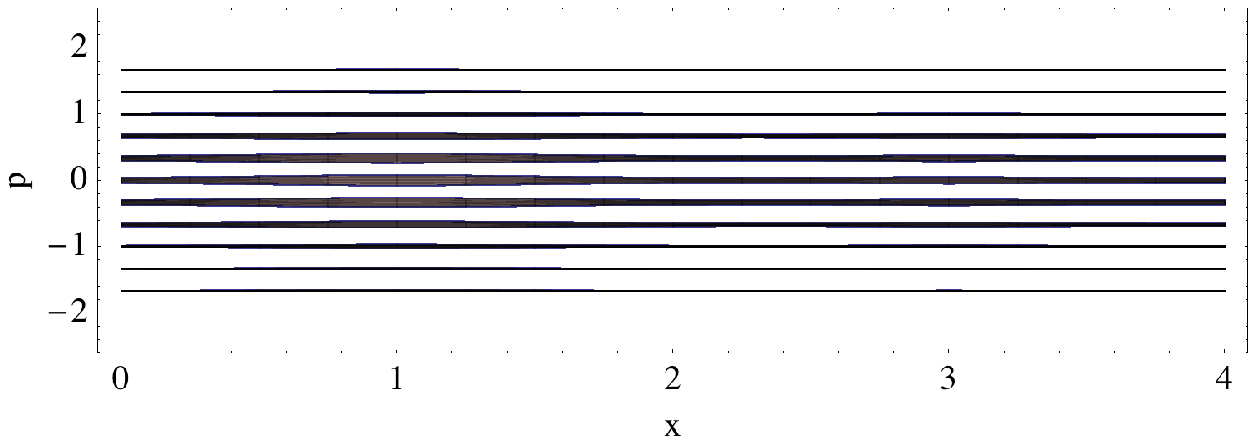}
\includegraphics[width=10cm]{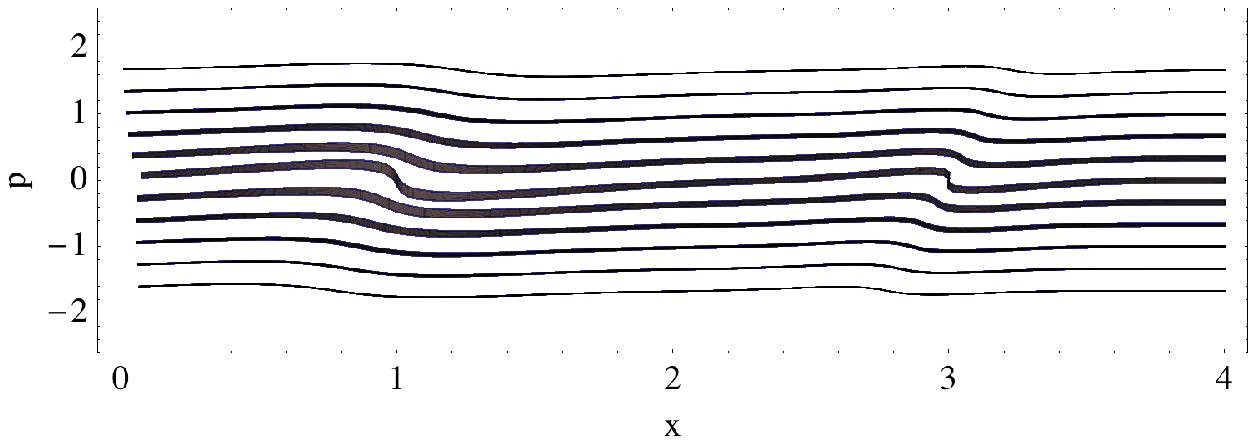}
\caption{Sketch of 1D phase--space evolution. The top panel shows flows with initially no
velocity gradients but density fluctuations (illustrated by the thickness variation of the lines). The bottom panel shows how the flows develop velocity gradients at later times. They can ultimately form multi--flow regions after they experience shell--crossing, in a way similar to what happens to dark matter flows. It is expected to happen preferably to flows with low initial velocities. Note that a flow with no initial velocity would behave
exactly like a cold dark matter component.}
\label{FlotNuInit}
\end{center}
\end{figure}

In a multi--fluid approach, the overall distribution function is obtained taking several fluids into account (see Fig. \ref{FlotNuInit} for illustration purpose).
More precisely, the overall distribution function $f^{\rm{tot}}$ has to be reconstructed from the single flows labeled by
$\vPi_{i}$,
\begin{equation}
 f^{\rm{tot}}(\eta,x^{i},p_{i})=\sum_{\vPi_{i}}{f^{\rm{one-flow}}(\eta,x^{i},p_{i};\vPi_{i})}=\sum_{\vPi_{i}}{n_{c}(\eta,\vx; \vPi_{i}) \delta_{\rm{D}}(p_{i}-P_{i}(\eta,\vx; \vPi_{i}))}.
\end{equation}
In the continuous limit, we thus have
\begin{equation}
 f^{\rm{tot}}(\eta,x^{i},p_{i})
 =\int{\dd^{3}\vPi_i}\,n_{c}(\eta,\vx;\vPi_{i}) \delta_{\rm{D}}(p_{i}-P_{i}(\eta,\vx;\vPi_i)),
\end{equation}
the parameter $\vPi_{i}$ being assumed to describe a 3D continuous field.

It means in particular that the momentum integrations in phase--space used in the standard description (i.e. for a single multi--flow fluid) to compute global physical quantities should be replaced in our description by a sum over the $\vPi_{i}$--fluids (i.e. a sum over all the possible initial momenta or velocities),
\begin{equation}\label{mapping1}
\int{\dd^{3}p_{i}}  \,f^{\rm{tot}}(\eta,x^{i},p_{i})\ \mF(p_{i})=  \int{\dd^{3} \vPi_{i}}\,n_c(\eta,\vx;\vPi_{i})\ \mF(P_{i}(\eta,\vx;\vPi_{i}))
\end{equation}
or equivalently
\begin{equation}
\int{\dd^{3}q^{i}} \,\frac{f^{\rm{tot}}(\eta,x^{i},q^{i})}{a^{3}}\ \mF(p_{i})=  \int{\dd^{3} \vPi_{i}}\,n(\eta,\vx;\vPi_{i})\ \mF(P_{i}(\eta,\vx;\vPi_{i}))
\label{mapping}
\end{equation}
for any function $\mF$. For each flow, the evolution equations are known but we still have to set the initial conditions to be able to use them in practice, see \ref{Initial conditions}. Before doing this, we will compute the multipole energy distribution associated with our description.

\subsection{The multipole energy distribution in the linear regime}

Of particular interest to compare our results to those of the Boltzmann approach is the computation of the overall
multipole energy distribution. We will focus on the total energy density ${\rho_{\nu}}$, the total energy flux dipole 
$\theta_{\nu}$ and the total shear stress $\sigma_{\nu}$. These quantities are directly related to the phase--space 
distribution function thanks  to the following relations (see \cite{Ma:1995ey}),
\begin{eqnarray}
{\rho_{\nu}}&=&-{T^0_{\phantom{0}0}}=\int{\dd^3{q^i}\,\dfrac{\epsilon(q^i)}{a^3}}  f,\\
\left({\rho_{\nu}}^{(0)}+{P_{\nu}^{(0)}}\right)\theta_{\nu}&=&
\ii k^i \Delta T^0_{\phantom{0}i}=\ii \Delta\left[\int{\dd^3{q^i}\,\dfrac{k^j q^j }{a^4}} f\right],\\
\left({\rho_{\nu}}^{(0)}+{P_{\nu}}^{(0)}\right)\sigma_{\nu}&=&-\left(\dfrac{k^i k^j}{k^2}-\dfrac{1}{3}\delta_{ij}\right)\left(\Delta T^i_{\phantom{0}j}-\dfrac{1}{3}\delta^i_j \Delta T^k_{\phantom{0}k}\right)
\nonumber \\&=&
\dfrac{1}{3}\Delta \left[\int{\dd^3{q^i}\,\dfrac{q^{i}q^{j}}{a^5 \epsilon(q^i)} \left(\delta^{ij}-3\frac{k^{i}k^{j}}{k^{2}}\right)}\,f\right],
\end{eqnarray}
where  $\rho_{\nu}^{(0)}$ and $P_{\nu}^{(0)}$ are the density and pressure of the neutrino fluid at background level and $\Delta$ stands for the perturbed part of the quantity it precedes. 

At linear order these quantities can be expressed with the help of the linear fields introduced in our description. More precisely, Eq. (\ref{mapping}) yields
\begin{align}
\rho_{\nu}^{(1)}&=4\vPi\int{\vPi^2 \dd{\vPi}\int_{-1}^1{\dd \mu \rho^{(1)}(\vPi,\mu)}}\label{monop},\\
\left({\rho_{\nu}}^{(0)}+{P_{\nu}^{(0)}}\right)\theta_{\nu}^{(1)}&=4\vPi \ii \int{\vPi^2 \dd{\vPi}\int_{-1}^1{\dd \mu \left[\rho^{(1)}(\vPi,\mu)\iniv(\vPi,\mu)\mu k+\rho_{\nu}^{(0)}(\vPi,\mu)k^i {V^i}^{(1)}(\vPi,\mu)\right]}}\label{dip},\\
\left({\rho_{\nu}}^{(0)}+{P_{\nu}}^{(0)}\right)\sigma_{\nu}^{(1)}&=-4\vPi  \int{\vPi^2 \dd{\vPi}\int_{-1}^1{\dd \mu \left[\rho^{(1)}(\vPi,\mu)\iniv^2(\vPi,\mu)\left(\mu^2-\dfrac{1}{3}\right)\right]}}\nonumber\\-8\vPi&  \int{\vPi^2 \dd{\vPi}\int_{-1}^1{\dd \mu \left[\rho^{(0)}(\vPi,\mu)\iniv(\vPi,\mu)\left(\dfrac{\mu k^i {V^i}^{(1)}(\vPi,\mu)}{k}-\dfrac{V^{(1)}(\vPi,\mu)}{3}\right)\right]}}.\label{quad}
\end{align}
For explicit calculation, note that $\rho^{(1)}(\vPi,\mu)=n^{(1)}(\vPi,\mu) \epsilon^{(0)}(\vPi,\mu)+n^{(0)}(\vPi,\mu) \epsilon^{(1)}(\vPi,\mu)$, where
\begin{equation}
{{\epsilon}^{(1)}(\vPi,\mu)}=\dfrac{m\iniv^{2}(\vPi,\mu)}{\sqrt{1-\iniv^{2}(\vPi,\mu)}}\phi-\frac{\ii\mu \iniv(\vPi,\mu)}{ak}\theta_{\tP}(\vPi,\mu)
\end{equation}
and that
\begin{equation}\label{velocity perturbation}
{V^i}^{(1)}(\vPi,\mu)=(1-\iniv^{2}) \iniv^{i}\phi-\frac{\ii}{k^{2}}\frac{\sqrt{1-\iniv^{2}}}{ma}\left(k^{i}
-k^{j}\iniv^{j}\iniv^{i} \right)\theta_{\tP}(\vPi,\mu).
\end{equation}
The physical quantities $\rho_{\nu}^{(1)}$, $\theta_{\nu}^{(1)}$ and $\sigma_{\nu}^{(1)}$ are source terms generating the metric fluctuations involved in the growth of the large--scale structure of the universe. We use them to compare the predictions of the
multi--fluid approach with those of the standard Boltzmann approach in Sect. \ref{numerical integration}.

\subsection{Initial conditions}\label{Initial conditions}

The initial time $\eta_{\rm{in}}$ is chosen so that the neutrino decoupling occurs at a time $\eta<\eta_{\rm{in}}$ 
and neutrinos become non--relativistic at a time $\eta>\eta_{\rm{in}}$. The initial conditions depend obviously on the cosmological model adopted. In this paper, we describe solutions corresponding to adiabatic initial conditions. It imposes to the quantities $\theta_\nu^{(1)}$ and $\sigma_\nu^{(1)}$, defined by Eqs. (\ref{dip}-\ref{quad}), to be zero but we still have some freedom in the way we assign each neutrino to one flow or to another. The initial conditions we present in the following correspond to a simple choice respecting the adiabaticity constraint.

\subsubsection{Initial momentum field $P_{i}(\etain,\vx;\vPi_{i})$}

The description we adopt is the following:
at initial time we assign to the flow
labeled by $\tau_i$ all the neutrinos whose momentum $P_{i}$ is equal to $\vPi_{i}$ within $\dd^{3}\vPi_{i}$. It obviously imposes
\begin{equation}
P_{i}(\etain,\vx;\vPi_{i})=\vPi_i.
\end{equation}
It implies in particular that ${P_{i}}^{(1)}(\vx,\etain;\vPi_{i})=0$ and consequently that $\theta_P(\vx,\etain;\vPi_{i})=0$.

\subsubsection{Initial number density field $n(\etain,\vx;\vPi_{i})$}
Although the velocity fields are initially uniform, it is not the case of the individual numerical density fields as we expect the total numerical density field to depend on space coordinates at linear order. 

Initial number densities are obviously strongly related to initial distribution functions in phase--space $f$. Before decoupling, the background distribution of neutrinos is expected to follow a Fermi--Dirac law $f_0$ with a temperature $T$ and no chemical potential (see e.g. Refs. \cite{2013neco.book.....L,Lesgourgues2006,1994ApJ...429...22M,Ma:1995ey} for a physical justification of this assumption),
\begin{equation}
f_0 \left(q\right)
\propto \frac{1}{1+\exp\left[q/(a k_B T)\right]},
\end{equation}
where $k_B$ is the Boltzmann constant and $q$ is the norm - previously defined, see the geodesic equation (\ref{deriv_pi}) - of the phase--space variable $q^i$.
As explained in Ref. \cite{2013neco.book.....L}, after neutrino decoupling, the phase--space distribution function of relativistic neutrinos is still a Fermi--Dirac distribution but modified by local fluctuations of temperature, whence
\begin{equation}\label{fform}
f\left(\etain,\vx,q\right)\propto \frac{1}{1+\exp\left[q/(a k_B (T+\delta T(\etain,\vx))\right]}.
\end{equation}
In terms of the variable $p_{i}$, it can be rewritten
\begin{equation}\label{fform2}
f\left(\etain,\vx,p_{i}\right)\propto \frac{1}{1+\exp\left[p(1+\phi(\vx, \etain))/(a k_B (T+\delta T(\vx, \etain)))\right]}.
\end{equation}
Given Eqs.  (\ref{comoving number density}) and (\ref{nncrelation}) and recalling that $f$ is non zero only for $p_i=\tau_i$ at initial time,
 the initial numerical density contrast follows directly,
\begin{equation}
\delta_n(\etain,x^{i};\vPi_i)=\dfrac{f^{(1)}(\etain,x^{i},\tau_{i})}{f^{(0)}(\etain,x^{i},\tau_{i})}+3\phi(\etain,x^i).
\end{equation}
The expression of $f^{(1)}(\eta,x^{i},p_{i})$ can be easily computed. It reads 
\begin{equation}
f^{(1)}(\etain,x^{i},p)=\frac{p}{a k_{B} T}\left(\frac{\delta T(\etain,x^i)}{T}-\phi(\etain,x^{i})\right)\frac{\exp[p/(a k_B T)]}{1+\exp[p/(a k_B T)]}\,f_{0}(p),
\end{equation}
which can be reexpressed in the form,
\begin{equation}
f^{(1)}(\etain,x^{i},p)=-\left(\frac{\delta T(\etain,x^{i})}{T}-\phi(\etain,x^i)\right)\frac{\dd f_{0}(p)}{\dd\log p}.
\end{equation}
The last step of the calculation consists in relating the local initial temperature fluctuations to the metric fluctuations  for adiabatic modes. It is a standard result,
which reads
${\delta T(\vx,\etain)}/{T(\etain)}=-{\psi(\vx,\etain)}/{2}$ (see e.g. \cite{Ma:1995ey}). We finally get the following expression for the linearized initial number density fluctuations,
\begin{equation}\label{initial delta}
\delta_{n}(\etain,\vx;\vPi_{i})=3\phi(\vx, \etain)+\left(\frac{\psi( \eta_{\rm{in}},\vx)}{2}+\phi(\etain,\vx)\right)\frac{\dd\log f_{0}(\vPi)}{\dd\log \vPi}.
\end{equation}
It can then easily be checked from Eqs. (\ref{monop}-\ref{quad}) that $\theta_{\nu}$ and $\sigma_{\nu}$ both vanish at initial time with this
choice of initial conditions.

\subsubsection{Early--time behavior}

A remarkable property of the initial fields we just computed is that they are isotropic, i.e. they do not depend on $\mu$. We know however that neutrinos develop an anisotropic
pressure which is a source term of the Einstein equations.
For practical purpose, e.g. to implement these calculations in a numerical code, it is therefore useful to examine in more
detail the sub-leading behavior at initial time. To that aim, we study how higher order multipoles arise at early time by decomposing the $\mu$ dependence of the fields $\delta_{n}$ and $\theta_{\tP}$ into 
Legendre polynomials,
\begin{equation}
\delta_{n}(\eta,\vx;\vPi,\mu)=\sum_{\ell}\delta_{n,\ell}(\eta,\vx;\vPi)\ (-\ii)^{\ell}P_{\ell}(\mu)
\end{equation}
and
\begin{equation}
\theta_{\tP}(\eta,\vx;\vPi,\mu)=\sum_{\ell}\theta_{\tP,\ell}(\eta,\vx;\vPi)\ (-\ii)^{\ell}P_{\ell}(\mu).
\end{equation}
In order to properly compute the source terms of the Einstein equations, one needs to know the expression of the number density multipoles
up to $\ell=2$ and the one of the momentum divergence multipoles up to $\ell=1$. The leading order behavior corresponding to these terms can be obtained easily from the motion equations (\ref{nFourier2}-\ref{thetaFourier2})
noting  that $\phi$ and $\psi$ are constant, $\mH$ scales like $1/a$ and $\sqrt{1-\iniv^{2}}$ scales like $a$ at superhorizon 
scales for adiabatic initial conditions.

Once the equations of motion are decomposed into Legendre polynomials, one gets successively,
\begin{eqnarray}
\theta_{\tP,0}&=&\frac{a}{\mH}\frac{m}{\sqrt{1-\iniv ^{2}}}k ^{2}\left(\phi+\psi\right)\label{inittheta0}\\
\theta_{\tP,1}&=&\frac{k}{2\mH}\ \theta_{\tP,0}\label{inittheta1}\\
\delta_{n,0}&=&3\phi+\left(\frac{\psi}{2}+\phi\right)\frac{\dd\log f_{0}(\vPi)}{\dd\log \vPi}\label{initdelta0}\\
\delta_{n,1}&=&\frac{k}{\mH}\left[\delta_{n,0}-\psi+2\phi)\right]\label{initdelta1}\\
\delta_{n,2}&=&-\frac{k}{3\mH}\delta_{n,1}-(1-\iniv^{2})^{1/2}\frac{\theta_{\tP,0}}{3am\mH}\label{initdelta2},
\end{eqnarray}
where $\theta_{\tP,\ell}$ scales like $a^{\ell+1}$ and $\delta_{n,\ell}$ scales like $a^{\ell}$ at leading order.

\subsection{Numerical integration}\label{numerical integration}
\label{Numerics}

This section aims at describing the numerical integration scheme developed to deal with a multi--fluid desciption and to compare its efficiency with that of the standard integration of the Boltzmann hierarchy.

\subsubsection{Method}

The equations of motion in Fourier space (\ref{nFourier2}) and (\ref{thetaFourier2}) 
are numerically integrated with the help
of a Mathematica program in which the time evolution of the metric perturbations is given. It is as usual determined by
the Einstein equations but in practice simply extracted from a
standard Boltzmann code (the code presented in \cite{2011ascl.soft09009P}). The initial conditions we implement correspond to the adiabatic expressions appearing in Eqs. (\ref{inittheta0}-\ref{initdelta2}). 
The main objective of this numerical experiment is to  check that the multipole energy distributions are identical 
when computed from the resolution of the Boltzmann hierarchy or from the equations of the multi--fluid description. 
In appendix \ref{Boltzmann hier}, we succinctly review the construction of the Boltzmann hierarchy. In this approach, energy multipoles are computed thanks to Eqs. (\ref{Boltzmann Hier0}-\ref{QuadHier}). The angular dependence of $q^i$ is taken into account via the Legendre polynomials used to decompose the distribution function in phase--space. Of course, because of the integration on $\dd^3 {q^i}$ necessary to compute the multipoles, the amplitude of $q$ has to be discretized for numerical integration.

In the  multi--fluid description, integrals that appear in the distribution energy (\ref{monop}-\ref{quad}) also involve a discretization on momentum directions, i.e. a discretization on $\mu$.
In both approaches, all integrals are estimated using the
 third degree Newton--Cotes formula (Boole's rule) which consists in approximating $\int_{x_{1}}^{x_{5}}\dd x $ by,
\begin{align}
\int_{x_1}^{x_5}{f(x)\dd{x}}&\approx 
\dfrac{2h}{45}[7 f(x_1)+32f(x_2)+12f(x_3)+32 f(x_4)+7f(x_5)],
\label{Newton--Cotes}
\end{align}
that is to say in using five discrete values regularly spaced, i.e. $x_{1+n}=x_1+n h$, $h=(x_{5}-x_{1})/4$, to compute the integral. In such a scheme, which gives exact results when integrating polynomials of order less than 6, the error term is proportional to $h^7$. In practice, we divide the $\vPi$, $q$ and $\mu$ ranges into respectively $N_\vPi$, $N_q$ and $N_\mu$ intervals - where $N_\vPi$, $N_q$ and $N_\mu$ are multiples of four - and we apply the integration scheme $N_\vPi/4$, $N_q/4$ and $N_\mu/4$ times. Besides, since 
the Fourier modes computed for $\mu$ and $-\mu$ are conjugate complex numbers when the initial gravitational potentials are real, we can restrict
our calculations to the range $[0,1]$ for $\mu$.
In the following, all the calculations are made using the WMAP5 cosmological parameters and the convergence tests are made for a neutrino mass of about 0.05 eV and for a wavenumber $k=0.2 h/$Mpc. 

\subsubsection{Results}

\begin{figure}[!h]
\begin{center}
\includegraphics[width=7cm]{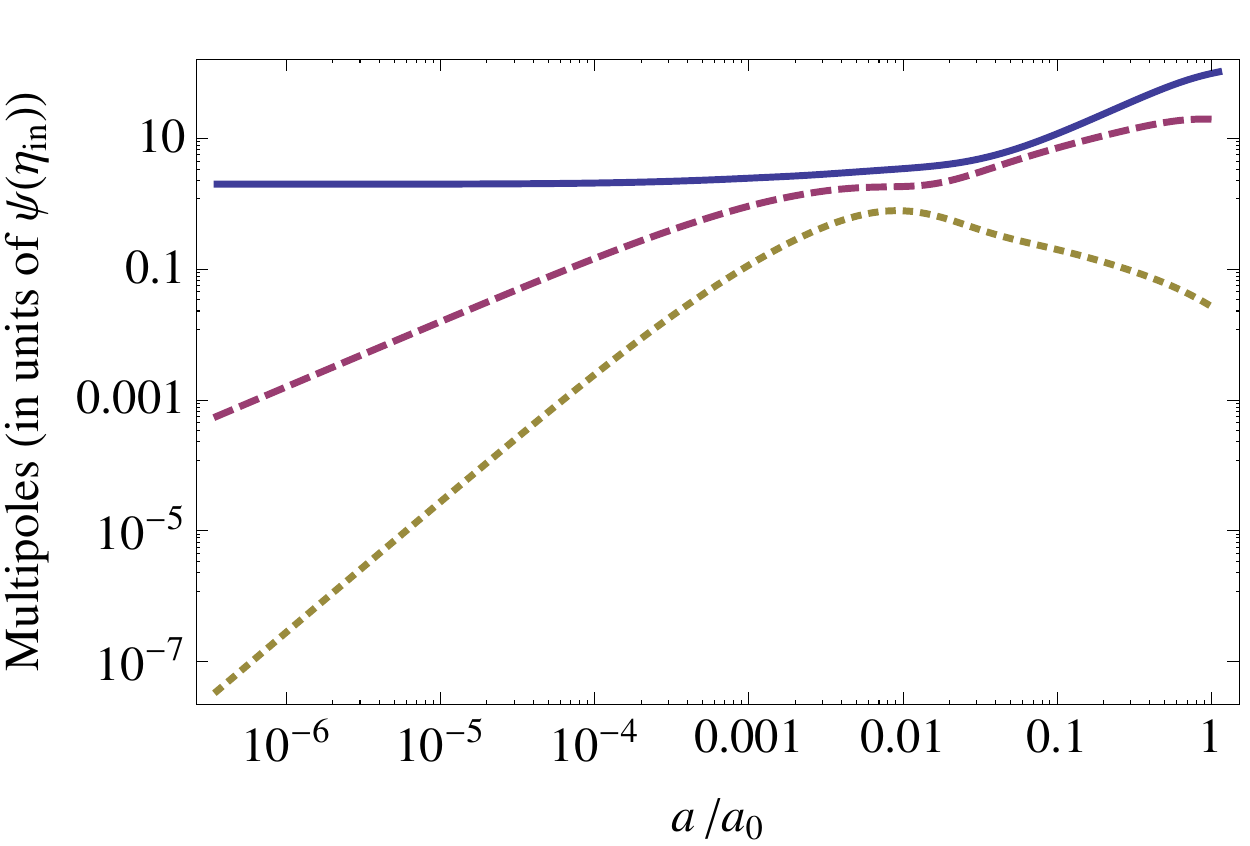}
\hspace{.1cm}
\includegraphics[width=7cm]{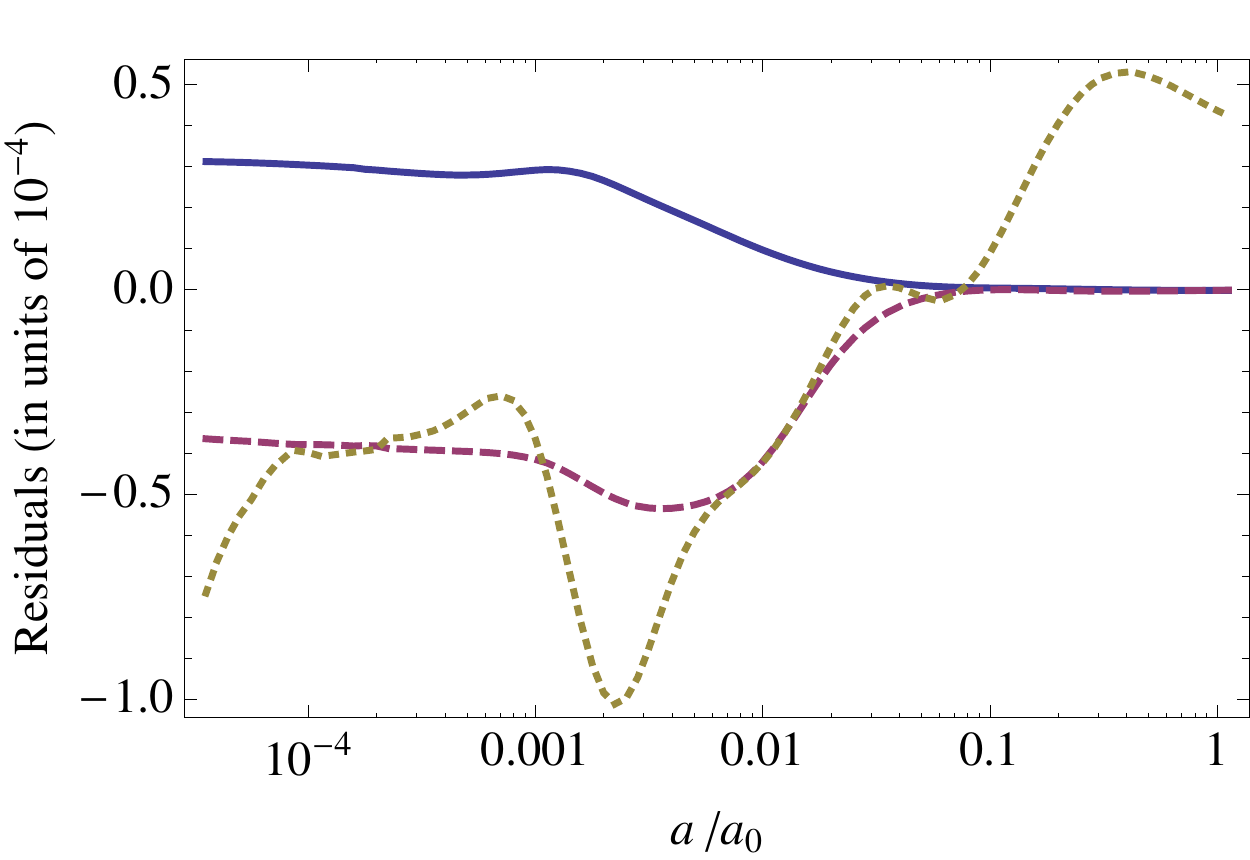}
\caption{Time evolution of the energy density contrast (solid line), velocity divergence (dashed line) and shear stress (dotted line) of the neutrinos. Left panel: the quantities are computed with the multi--fluid approach. 
Right panel:  residuals (defined as the relative differences) when the two methods are compared. Numerical integration has been
done with 100 values of $\vPi$ and $q$. The resulting relative differences are of the order of a fraction of $10^{-4}$.}
\label{multipoles}
\end{center}
\end{figure}

We first compare the consistency of the two approaches by varying $N_\vPi$ and $N_\mu$ on one side and $N_q$ and $\ell_{\rm max}$ on the other side, where $\ell_{\rm max}$ is the order at which the Boltzmann hierarchy is truncated. 
As illustrated  on Fig. \ref{multipoles}, the values computed in both descriptions can reach
an extremely good agreement when these parameters are large enough: the results correspond to $N_{\mu}=12$, $\ell_{\rm max}=6$ and $N_{q}=N_{\vPi}=100$ and the accuracy is better than $10^{-4}$.
Performing several tests, we realized  that the main limitation in the relative precision is the number of points we put in the $\vPi$ or $q$ intervals. With 16 values for each, only percent accuracy is reached (see Table \ref{relative errors} for more details regarding the relative precision one can get). Meanwhile, the parameters $N_{\mu}$ and $\ell_{\rm max}$ do not appear as critical limiting factors in the accuracy of the numerical integration, provided of course that they are not too small. For instance, with 
$N_{\mu}=12$, the numerical scheme (\ref{Newton--Cotes}) allows to reach an  exquisite accuracy and with $N_{\mu}=8$,
it is still possible to reach $10^{-3}$. These tests show that the extra cost of the use of our representation, where neutrinos are described by a set of 
$2\times N_{\mu}\times N_{\vPi}$ equations instead of $\ell_{\rm max}\times N_{q}$, is not dramatically large.

\begin{table}[!h]
\begin{center}
\begin{tabular}{|c|ccc|}
\hline
$N_q$ ($\downarrow$) and $N_\vPi$($\rightarrow$)&{16}&{40}&{100}\\
\hline
16&$10^{-2}$&$5\,10^{-3}$&$5\,10^{-3}$
\\
40&$10^{-2}$&$10^{-3}$&$2\,10^{-4}$
\\
100&$10^{-2}$&$10^{-3}$&$10^{-4}$\\
\hline
\end{tabular}
\end{center}
\caption{Relative errors (averaged in time between $a/a_{0}=10^{-5}$ and  $a/a_{0}=1$) between the results obtained with the Boltzmann hierarchy and those obtained with the multi--fluid approach. For each value of $N_q$ and $N_\vPi$,  the largest magnitude of the relative error on either the density, the dipole or the shear is given. Calculations are made with $N_{\mu}=12$ and $\ell_{\rm max}=6$.}
\label{relative errors}
\end{table}

\medskip

\begin{figure}[!h]
\begin{center}
\includegraphics[width=7cm]{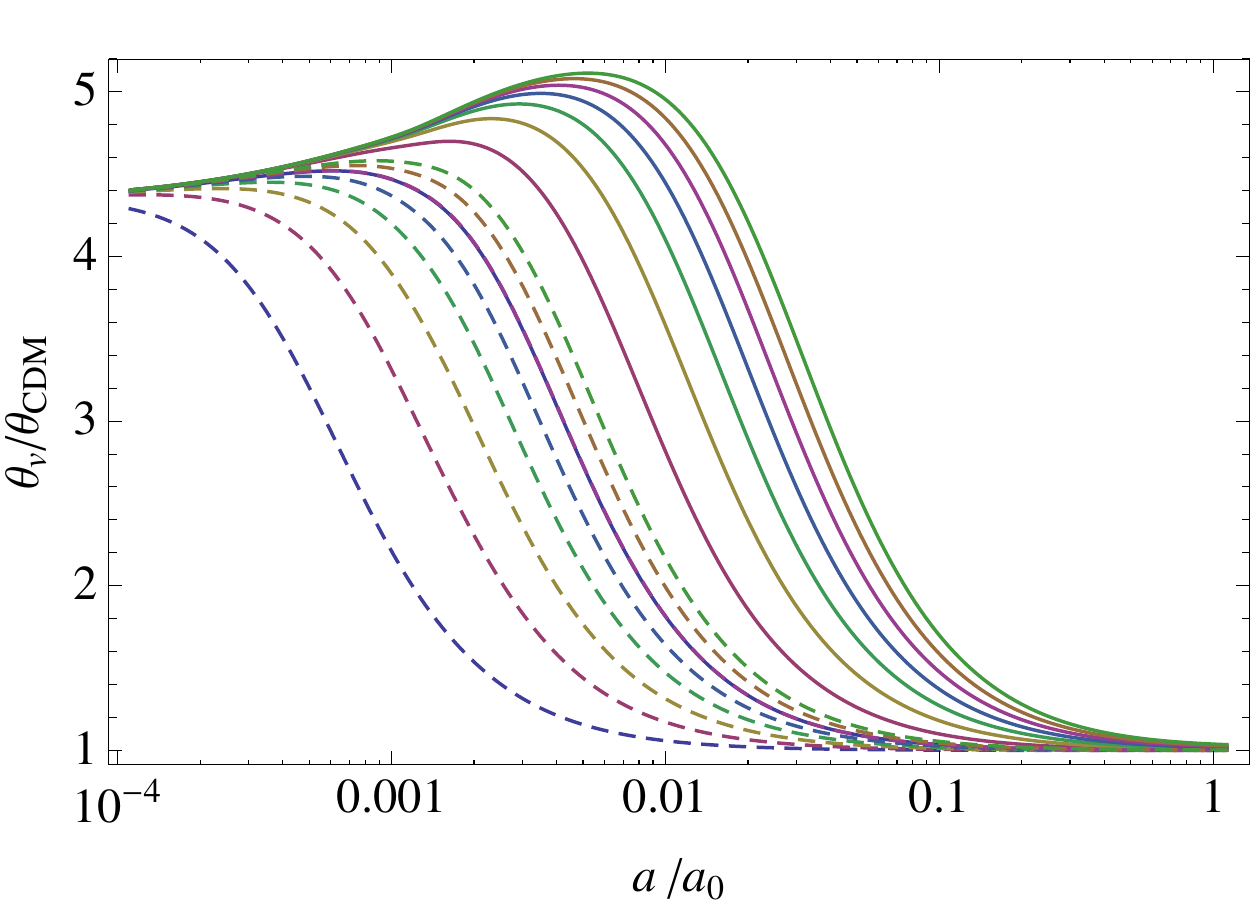}
\hspace{.2cm}
\includegraphics[width=7cm]{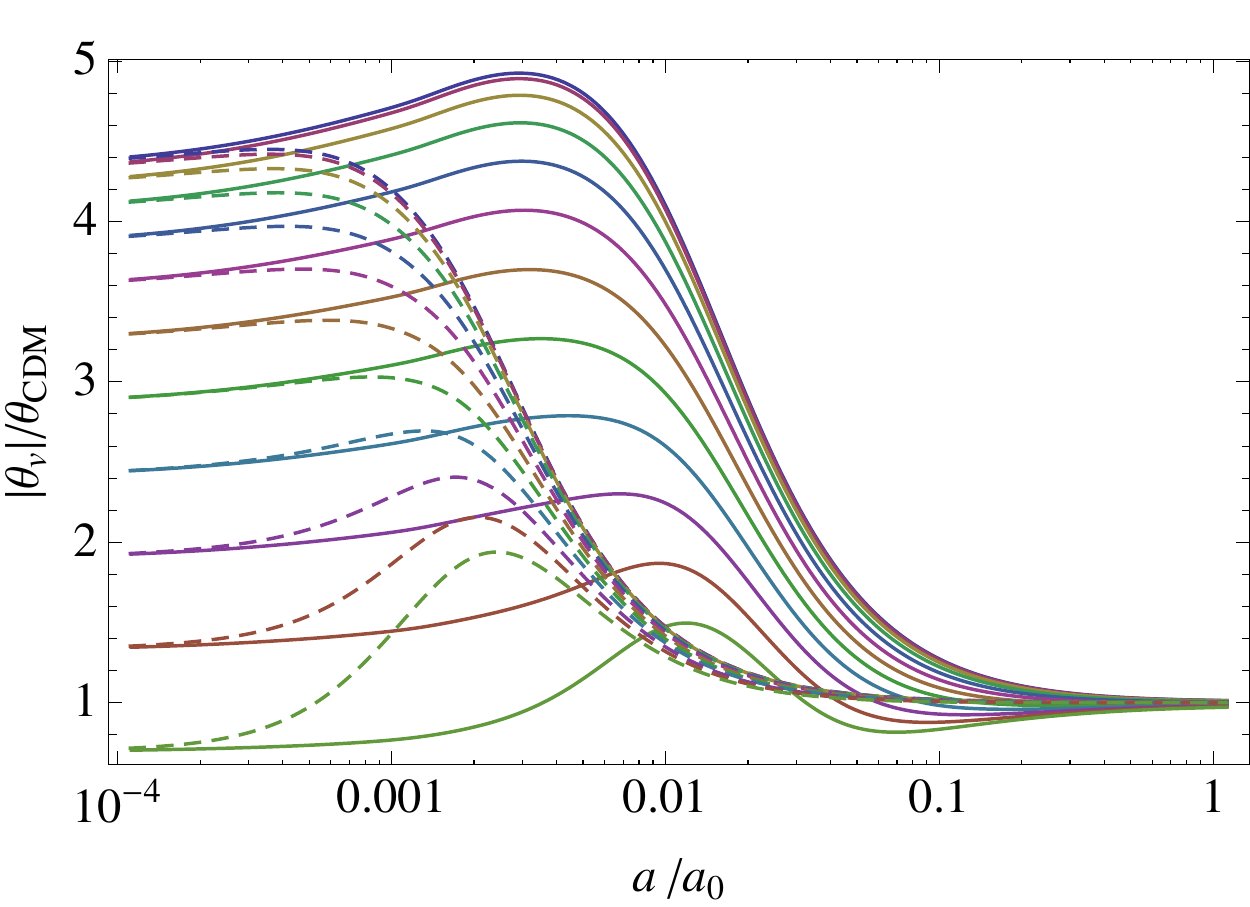}
\caption{Time evolution of the velocity divergence. Left panel: values of $\vPi$ range from $0.86\,k_{B}T_{0}$ (bottom lines) to $7\,k_{B}T_{0}$ (top lines) with $\mu=0$. Right panel is for $\vPi=3.5\,k_{B}T_{0}$ and  $\mu$ ranging from $\mu=0$ (top lines) to $\mu=1$ (bottom lines). The time evolution of the velocity divergence of each flow is plotted in units of the dark matter velocity divergence. The solid lines are for a $0.05$ eV neutrino and the dashed lines for a $0.3$ eV neutrino.}
\label{VelDivConvergence}
\end{center}
\end{figure}

\begin{figure}[!h]
\begin{center}
\includegraphics[width=7cm]{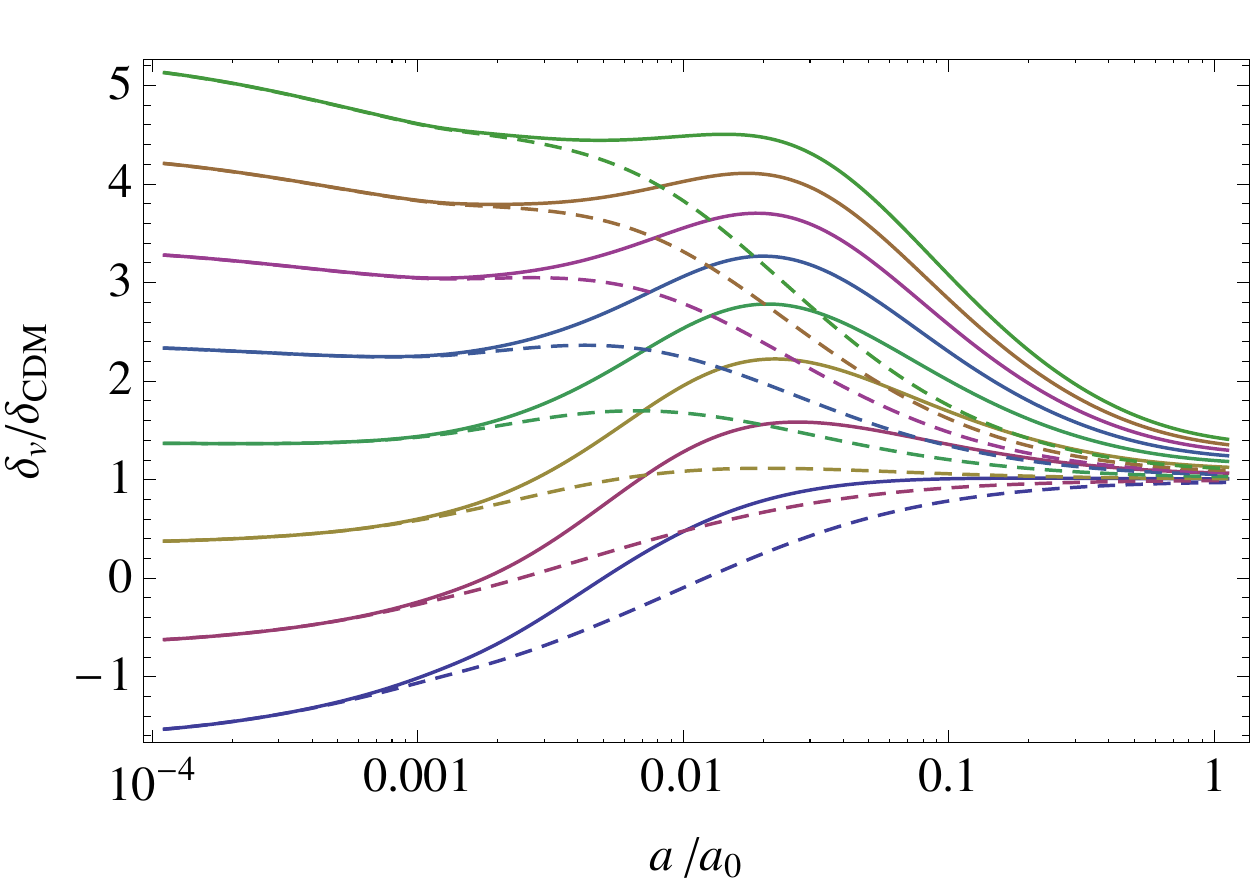}
\hspace{.2cm}
\includegraphics[width=7cm]{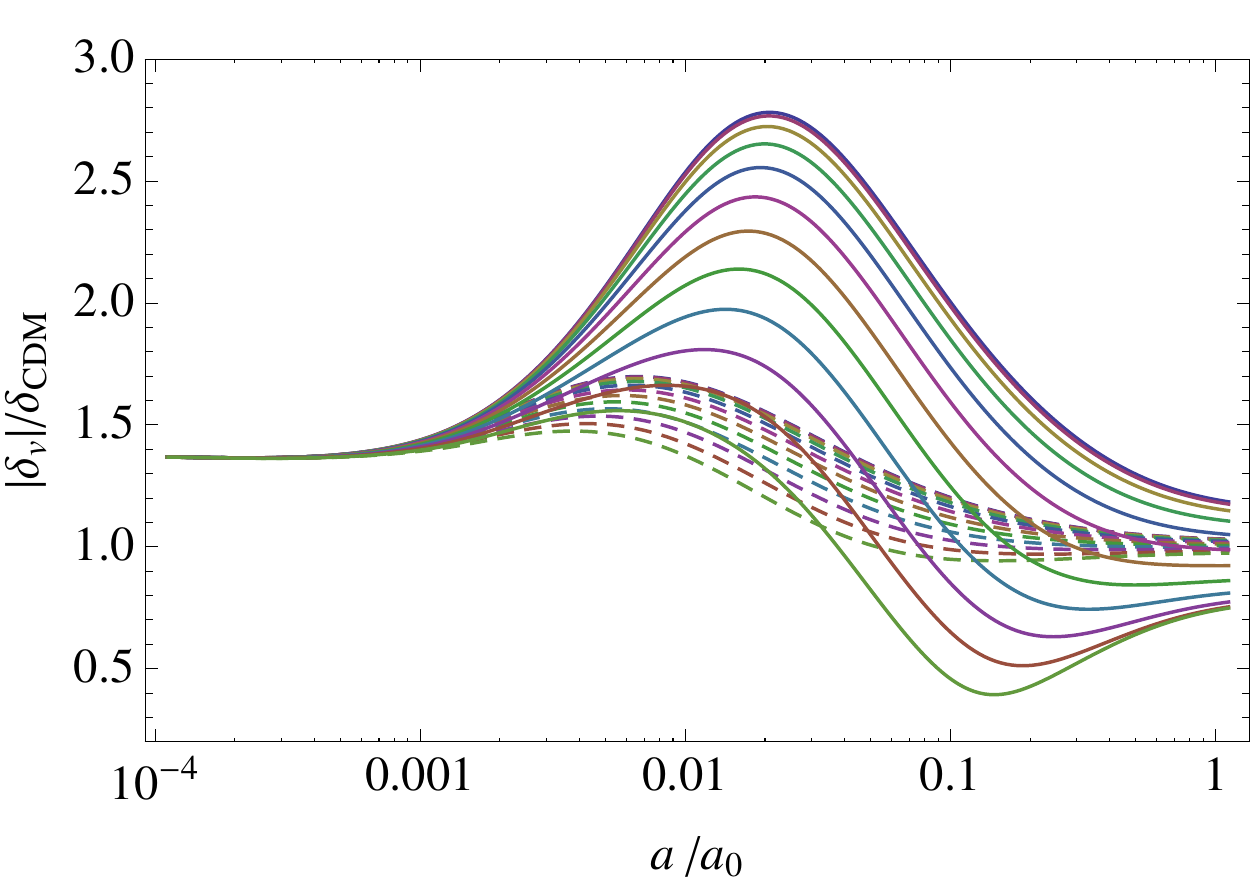}
\caption{Same as the previous plot for the number density contrast.}
\label{DeltaNConvergence}
\end{center}
\end{figure}

To finish, we illustrate the fact that the multi--fluid approach, by its specificity, allows to show the convergence of  the number 
density contrast and of the velocity divergence of each flow  to the ones of the Cold Dark Matter (CDM) component. 
Each flow is characterized  by two parameters: its initial momentum modulus $\vPi$ 
and the angle $\mu$ between its initial velocity vector $\iniv^{i}$ and the wave vector $k^{i}$. 
Unsurprisingly, for a fixed mass, 
the smaller the initial momentum $\vPi$ is,  the larger its decay rate is and, for a fixed momentum, this rate increases when the neutrino mass increases. 
This is illustrated on the left panels of
Figs. \ref{VelDivConvergence} and \ref{DeltaNConvergence}, which show the convergence of the amplitudes of the fluctuations of several neutrino flows to the fluctuations of the dark matter component when $\mu$ is set to zero\footnote{The velocity divergence that appears on Fig. \ref{VelDivConvergence} is related to the momentum divergence by Eq. (\ref{velocity perturbation}).}. 
Besides, the right panels of these figures show that 
this convergence is modulated by the value of $\mu$. It is all the more rapid that $\mu$ is close to $1$, that is when the velocity is along
the wave mode.
It should be noted however that these plots
only partially describe the settling of the flows in the dark matter component as they give only the absolute values of complex
Fourier modes. When $\mu$ is not zero, the fluctuations of the neutrino flows and of the CDM component are indeed expected to be out of phase for a while.
A last remark is the observation that the convergence of the velocity divergence is more rapid than the one of
the number density. It simply illustrates the fact that the former acts as a source term of the latter in the motion equations.

\section{Conclusions}
We have developed an alternative approach 
to the method based on the Boltzmann hierarchy to account for massive neutrinos in non--linear cosmological calculations. In this new description, neutrinos are treated as a collection of single--flow fluids and their behavior is encoded in fluid equations derived from conservation laws
or from the evolution of the phase--space distribution function. The resulting fluid equations, (\ref{neom}) and (\ref{Peom}), are derived at linear level with respect to the metric perturbations but at full non--linear level with respect to the density fluctuations and velocity divergences.
They can easily be compared to the equations resulting from the standard study of the distribution function of a single hot fluid of dark matter particles. 
After having considered in detail these equations in the linear regime, 
we have shown precisely how a proper choice of the single--flow fluids and a proper choice of the initial 
conditions allow to recover the physical behavior of the overall neutrino fluid.  These initial conditions are given explicitly in the case of initially adiabatic metric perturbations. 

We then check that the two descriptions are equivalent at linear level through numerical experiments. The conclusion is that
the whole macroscopic properties of the neutrino fluid can actually be accounted for by studying such a collection of flows with an arbitrary 
precision (in practice we reached a $10^{-5}$ relative precision). An additional information exists in our approach since it also describes the physics of each flow separately. We illustrate this point by showing how individual neutrino
flows converge to the CDM component as a function of their initial momentum and of the neutrino mass at play.

This representation opens the way to a genuine and fully non--linear treatment
of the neutrino fluid during the late stage of the large--scale structure growth as the two evolution
equations satisfied by each flow can be incorporated separately into
the equations describing the non--linear dynamics of this growth. In particular, it should be possible to apply resummation techniques such
as those introduced in \cite{2006PhRvD..73f3519C,2012PhRvD..85f3509B,2013PhRvD..87d3530B} or to 
incorporate the neutrino component at non--linear level in approaches such as
\cite{2008JCAP...10..036P,2008ApJ...674..617T}.
We leave for future work the examination of the importance
of non--linear effects on observables such as power spectra.
\\

\textbf{Acknowledgements}: The authors are grateful to Cyril Pitrou, Jean-Philippe Uzan, Pierre Fleury, Romain Teyssier 
and Atsushi Taruya for
insightful discussions and encouragements. FB also thanks the YITP of the university of Kyoto and the RESCUE of the university of 
Tokyo for hospitality during the completion of this manuscript. 
This work is partially supported by the grant ANR-12-BS05-0002 of the French Agence Nationale de la Recherche.

 \appendix
 \section{The Boltzmann hierarchies}\label{Boltzmann hier}

Starting from the Vlasov equation (\ref{Boltzmann}), one can build hierarchies to describe the evolution of the moments of the phase--space distribution function. There are several ways to do this.

\subsection{A Boltzmann hierarchy from tensor field expansion}

A first hierarchy can be built 
by integrating equation (\ref{Boltzmann}) with respect to $\dd^3{\mathbf{q}}$, weighted by products of $\dfrac{q^i}{a\epsilon}$. To that end, it is useful to introduce the tensorial fields $A$, $A^i$, $A^{ij}$,... defined as (see \cite{vanderijtphd})
\begin{eqnarray}
\label{A definition2}
A&\equiv&\rho,\\
A^{ij...k}&\equiv&\int{\dd^3{\mathbf{q}}\left[\dfrac{q^i}{a\epsilon}\dfrac{q^j}{a\epsilon} ... \dfrac{q^k}{a\epsilon}\right]\dfrac{\epsilon f}{a^3}}.
\end{eqnarray}
{After multiplying Eq. (\ref{Boltzmann}) by adequate factors such as $\epsilon/a^{3}$, $\epsilon/a^{3}\,q^{i}/(a\epsilon)$ and in general $\epsilon/a^{3}\,q^{i_{1}}/(a\epsilon) \dots
q^{i_{n}}/(a\epsilon)$, integrations by parts directly give the desired hierarchy of equations. For $A$ it leads to }
  
\begin{equation}\label{Aevol2}
\partialeta  A+(\mH-\partialeta \phi)(3A+A^{ii})+(1+\phi+\psi)\partial_{i}A^{i}+2A^{i}\partial_{i}(\psi-\phi)=0,
\end{equation}
for $A^{i}$ it leads to,
\begin{equation}\label{Aievol}
\partialeta {A^{i}}+4(\mH-\partialeta \phi)A^{i}+(1+\phi+\psi)\partial_{j}A^{ij}+A\partial_{i}\psi+A^{ij}\partial_{j}\psi
-3 A^{ij}\partial_{j}\phi+A^{jj}\partial_{i}\phi=0,
\end{equation}
and in general it leads to the following equation,
\begin{align}\label{Aijkevol}
\partialeta {A}^{i_{1}\dots i_{n}}+(\mH-\partialeta \phi)\left[(n+3)A^{i_{1}\dots i_{n}}-(n-1)A^{i_{1}\dots i_{n}jj}\right]
\nonumber\\
+\sum_{m=1}^{n}(\partial_{i_{m}}\psi) A^{i_{1}\dots i_{m-1}i_{m+1}\dots i_{n}}
+\sum_{m=1}^{n}(\partial_{i_{m}}\phi) A^{i_{1}\dots i_{m-1}i_{m+1}\dots i_{n}jj}
\nonumber\\
+(1+\phi+\psi)\partial_{j}A^{i_{1}\dots i_{n}j}
+\left[(2-n)\partial_{j}\psi-(2+n)\partial_{j}\phi
\right]A^{i_{1}\dots i_{n}j}=0
\end{align}
Note that this hierarchy of coupled equations retains the same level
of non--linearities as Eqs. (\ref{neom}) and (\ref{Peom}). Once linearized, it is equivalent to the standard hierarchy of equations describing the multipole
decomposition of the distribution function perturbation as given below.

 \subsection{A Boltzmann hierarchy from harmonic expansion}

 We recall here the standard construction of the Boltzmann hierarchy, i.e. of the hierarchy that Boltzmann codes usually implement (see Refs. \citep{1994ApJ...429...22M,Lesgourgues2006,Ma:1995ey,vanderijtphd} for more details).
 It is based on a decomposition of the phase--space distribution function $f(\vx,\vq)$ into a homogeneous part and an inhomogeneous contribution,
 \begin{equation}
f(\vx,\vq)=f_{0}(q)\left[1+\Psi(\vx,\vq)\right]
\end{equation}
and a decomposition of the latter into harmonic functions.
At linear order, the Vlasov equation for $f$ (\ref{Boltzmann}) leads to the following equation for $\Psi$,
\begin{equation}
\partialeta \Psi+\dfrac{q}{a\epsilon} \hn^{i}\partial_{i}\Psi+\frac{\dd \log f_{0}(q)}{\dd\log q}
\left(\partialeta \phi -\frac{a\epsilon}{q}
\hn^{i}\partial_{i}\psi\right)=0,
\end{equation}
where the local momentum is defined trough its norm $q$ and its direction $\hn$. 
 In momentum space, the only dependence on $\vk$ is through its angle with 
 $\hn$, so we define $\alpha \equiv \hk.\hn$ and rewrite the linearized Boltzmann equation as
\begin{equation}
\partialeta \tPsi+\ii\alpha k\frac{q}{a\epsilon} \tPsi+
\left(\partialeta \phi -\ii\alpha k\frac{a\epsilon}{q}\psi
\right)=0,\label{BoltztSpi}
\end{equation}
 where $\tPsi\equiv \left(\frac{\dd \log f_{0}(q)}{\dd\log q}\right)^{-1}\Psi$. The next step is to expand
 $\tPsi$  using Legendre polynomials thus we introduce the moments
 $\tPsi_{\ell}$,
 \begin{equation}
\tPsi=\sum_{\ell}(-\ii)^{\ell}\tPsi_{\ell}\,P_{\ell}(\alpha),
\end{equation}
where $P_{\ell}(\alpha)$ is the Legendre polynomial of order $\ell$. By plugging this expansion into the Boltzmann equation 
(\ref{BoltztSpi}), one obtains the standard hierarchy,
\begin{eqnarray}
\partialeta\tPsi_{0}(\eta,q)&=&-\frac{qk}{3a\epsilon}\tPsi_{1}(\eta,q)-\partialeta\phi(\eta)\label{Boltzmann Hier0}\\
\partialeta\tPsi_{1}(\eta,q)&=&\frac{qk}{a\epsilon}\left(\tPsi_{0}(\eta,q)-\frac{2}{5}\tPsi_{2}(\eta,q)\right)-\frac{a\epsilon k}{q}\psi(\eta),\label{Boltzmann Hier1}\\
\partialeta\tPsi_{\ell}(\eta,q)&=&\frac{qk}{a\epsilon}\left[
\frac{\ell}{2\ell-1} \tPsi_{\ell-1}(\eta,q)-\frac{\ell+1}{2\ell+3}\tPsi_{\ell+1}(\eta,q)
\right]\ \ (\ell\ge 2).\label{Boltzmann Hierl}
\end{eqnarray}

Since this hierarchy is infinite, it is of course necessary to truncate it at a given order for practical implementation.
Finally, relevant physical quantities can be built out of the coefficients $\tPsi_{\ell}(\eta,q)$,
\begin{eqnarray}
\rho_{\nu}^{(1)}(\eta)&=&4\vPi\int q^{2}\dd q \frac{\epsilon f_{0}(q)}{a^{3}}\frac{\dd \log f_{0}(q)}{\dd\log q}\,\tPsi_{0}(\eta,q)
\label{rhoHier}\\
(\rho_{\nu}^{(0)}+P_{\nu}^{(0)})\theta_{\nu}^{(1)}(\eta)&=&\dfrac{4\vPi}{3}\int q^{2}\dd q \frac{\epsilon f_{0}(q)}{a^{3}}\frac{\dd\log f_{0}(q)}{\dd\log q}\,\frac{q}{a\epsilon}\tPsi_{1}(\eta,q)
\label{DipHier}\\
(\rho_{\nu}^{(0)}+P_{\nu}^{(0)})\sigma_{\nu}^{(1)}(\eta)&=&\dfrac{8\vPi}{15}\int q^{2}\dd q \frac{\epsilon f_{0}(q)}{a^{3}}\frac{\dd\log f_{0}(q)}{\dd\log q}\,\left(\frac{q}{a\epsilon}\right)^{2 }\tPsi_{2}(\eta,q).
\label{QuadHier}
\end{eqnarray}
Note that as the numerical integration of the Boltzmann hierarchy gives access to $\tPsi_{\ell}$, expressions of
$\rho^{(1)}(\eta)$, $\theta_{\nu}^{(1)}$ and $\sigma_{\nu}^{(1)}$ are computed from Eqs. (\ref{rhoHier}-\ref{QuadHier}).

\bibliographystyle{JHEP}
\bibliography{neutrinos}
 \end{document}